\documentclass[twocolumn,pre,showpacs,floats,superscriptaddress]{revtex4}

\usepackage[dvips]{graphicx}
\usepackage{amssymb}
\usepackage{amsmath,bm}
\usepackage{psfrag}
\usepackage{epsfig}
\usepackage{float}

\begin{document}
\title{A longwave model for strongly anisotropic growth of a crystal step}

\author{Mikhail Khenner}
\affiliation{Department of Mathematics, Western Kentucky University, Bowling Green, KY 42101}

\begin{abstract}
A continuum model for the dynamics of a single step with the strongly anisotropic line energy is formulated and analyzed. The step grows by attachment of adatoms from the lower terrace,
onto which atoms adsorb from a vapor phase or from a molecular beam, and the desorption is non-negligible (the ``one-sided" model).  
Via a multi-scale expansion, we derived a longwave, strongly nonlinear, and strongly anisotropic evolution PDE for the step profile. 
Written in terms of the step slope, the PDE can be represented in the form similar to a convective Cahn-Hilliard equation.
We performed the linear stability analysis and computed the nonlinear dynamics. Linear stability depends on whether the stiffness is minimum or maximum in the direction of the step growth. It also depends nontrivially on the combination of the anisotropy strength parameter and the atomic flux from the terrace to the step. Computations show formation and coarsening of a hill-and-valley structure superimposed onto a large-wavelength profile, which independently coarsens. Coarsening laws for the hill-and-valley structure are computed for two
principal orientations of a maximum step stiffness, the increasing anisotropy strength, and the varying atomic flux. 

\end{abstract}

\date{\today}
\maketitle


\section{Introduction}
\label{Intro}

\begin{figure}[h]
\vspace{0.25cm}
\centering
\includegraphics[width=2.0in]{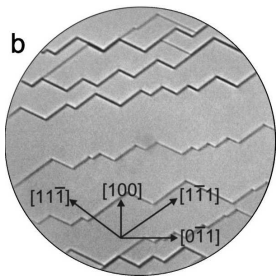}
\vspace{-2.5cm}
\caption{
Faceted step bunches (dark lines) and individual steps (light lines) on the Nb(011) surface. 
(Reprinted with permission from Ref. \cite{OSYP}. Copyright 2002, American Vacuum Society.)
}
\label{Experiments}
\end{figure}

In several well-known experimental papers it was observed that crystal steps on homoepitaxially growing, clean semiconductor or metal surfaces become faceted and form corners  \cite{OSYP}, \cite{OO}.
In Ref. \cite{OO} the train of monoatomic steps on Si(111) develops a zig-zag (sawtooth) in-phase instability. In Ref. \cite{OSYP} two-to-five monoatomic steps bunch on the surface of Nb(011), 
forming a multistep which then becomes faceted as it evolves, and some distinct monoatomic steps also facet. 
It also appears from the micrographs
(see Fig. \ref{Experiments}), that corners where the facets meet are very sharp, and the facets are remarkably straight and seemingly free of kinks. 
Although certainly the multi-step kinetics is very important in these experiments (see also Ref. \cite{XSZZWT}),
it is conceivable that equilibrium thermodynamics is at least partially responsible for step faceting and corner formation \cite{DPKM}. 

Guided by these examples, in this paper we model strongly anisotropic step dynamics for a single step that grows by attachment of adatoms from the lower terrace, 
onto which atoms adsorb from a vapor phase or from a molecular beam (and weakly adsorbed atoms de-adsorb back into the ambient).  
Weakly anisotropic step dynamics in this simple setup was modeled 
by Saito \& Uwaha \cite{SU}, who derived the Kuramoto-Sivashinsky type equation via the weakly nonlinear analysis near the instability threshold. (In Ref. \cite{MSOW} their work is extended to include the diffusion
on, and attachment of adatoms from, the upper terrace.)  
In the weakly anisotropic formulation the facets of a step are precluded from being straight lines, 
and the corners are smooth. The former is achieved by employing the twice-differentiable expression for the step line energy $\beta$ 
(thus step stiffness $\tilde \beta$ is defined for all orientations),
and the latter is achieved by using small values of the anisotropy strength parameter $\alpha$ (see Refs. \cite{S,LLRV} for the detailed review and discussion of the pertinent anisotropy model).

This paper is partially motivated by the fact that the analysis by Saito \& Uwaha, being to-date the only one published which targets the anisotropic step dynamics, does not reveal effects of the anisotropic line energy on the step linear stability; we try to close this gap 
for the case of strong anisotropy and
also analyze some aspects of the nonlinear step dynamics. By strong anisotropy we mean that through smooth, twice-differentiable
line energy the facet is still not allowed to have zero curvature, but sharp corners are possible - due to the assumed
negative values of the step stiffness for some orientations - 
when there is no material deposition on the surface and the temperature is very low or zero. (Absence of deposition and low temperature imply a vanishingly small or an insignificant density of kinks on the step - thus the step is stationary.) 
As the temperature increases (and so does the number of kinks, making the step rough), and the step starts to evolve,
the \textit{tendency} to form sharp corners remains due to assumed strong anisotropy, 
but the penalty (regularization) term included in the line energy imposes a radius of curvature and thus rounds up sharp corners before they form. 
The physical origin of the step corner 
regularization is thought to be the energy, $E\left(n,n_x\right)$, of kinks interaction inside a corner, much the same way the regularization is believed to emerge for a dynamic corner on a two-dimensional crystal surface 
(where interacting one-dimensional steps provide the needed contribution to the surface energy). Here $n(x)$ is the density of kinks, and $x$ is the direction across the corner \cite{Regular}.

%

\section{Problem formulation and the derivation of the strongly nonlinear evolution equation for the step}
\label{Formulation}

We consider morphological evolution of an unstable monoatomic step on a crystal surface. A step grows by the flux of adatoms from the lower terrace, and the line energy is assumed
anisotropic. (If the initially straight step is at $z=0$, then the lower terrace is the domain $z>0$.)

The governing equations of the model are the steady-state diffusion equation for the concentration of adatoms on the lower terrace, the mass conservation condition at the terrace edge
(the step),  the (modified) Gibbs-Thomson boundary condition for the concentration at the step \cite{GDN}, and the boundary condition for the concentration on the lower terrace far from the step:
\begin{equation}
D\nabla^2 C - \frac{C}{\tau} = - f,
\label{diff_eq}
\end{equation}
\begin{subequations}
\begin{eqnarray}
z=H(x,t):\quad V_n & \equiv & H_t \cos{\theta} = D\Omega\nabla C\cdot {\mathbf n}, \label{H_eq}\\
C &=& C_{eq}\left[ 1+ \frac{\Omega}{k_B\bar T}\left\{ \tilde \beta - \right.\right. \label{GT} \\
& & \hspace{-0.7cm}\left.\left.\qquad \qquad \bar \delta(|\alpha|)\left(\frac{\kappa^2}{2}+\frac{\kappa_{ss}}{\kappa}\right)\right\}\kappa \right], \nonumber
\end{eqnarray}
\end{subequations}
\begin{equation}
z\rightarrow \infty:\quad C = \tau f;
\label{infty}
\end{equation}
\[
{\mathbf n} = \left(-H_x \cos{\theta}, \cos{\theta}\right),\quad \cos{\theta} = \left(1+H_x^2\right)^{-1/2},
\]
\[
\bar \delta(|\alpha|)>0,\quad \tilde \beta =  \beta_0 (1-15\alpha \cos{4\theta}).
\]
In Eqs. (\ref{diff_eq}) - (\ref{infty}) $D$ is the diffusivity, $C$ the concentration of adatoms on a terrace, $\tau$ the desorption time, $f$ the molecular or atomic flux impinging on a terrace, $H(x,t)$ the step profile, $\theta$ the angle of the unit normal ${\mathbf n}$ to the step with the $z$-axis (the principal crystal direction), $V_n$ the normal velocity of the step, $\Omega$ the atomic volume, $C_{eq}$ the equilibrium concentration, $s$ the arclength
along the step, $k_B\bar T$ the Boltzmann factor, $\kappa$ the step curvature, $\beta_0$ the mean energy of a step line, $\tilde \beta$ the corresponding stiffness, $\alpha$ the anisotropy strength, and $\bar \delta$ the regularization parameter. Note that the expression in the curly brackets is the \textit{regularized} stiffness.

The formulation in Eqs. (\ref{diff_eq})-(\ref{infty}) differs from the one in Ref. \cite{SU} in two respects. 
\begin{figure}[h]
\vspace{-2.0cm}
\centering
\includegraphics[width=3.0in]{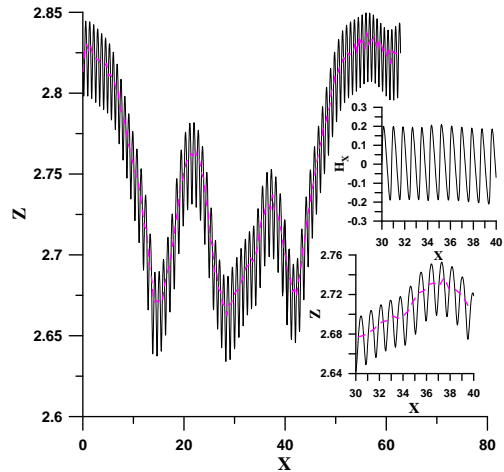}
\vspace{-1.5cm}
\caption{(Color online.) Late time step morphology. $\alpha=0.1$, $F=2$. Dashed line: large wavelength modulation of the mean step position. This line connects the grid nodes whose numbers are the average values of
all grid nodes on an uphill or a downhill - thus the locations of such grid nodes do not coincide with a midpoint on a slope, as can be seen in the bottom inset (but they are close). 
Bottom inset: zoom into the middle section of the main figure. Top inset: zoom into the graph of the step slope, $H_x$.  Note: the axes labels 
in the Figures are capitalized for visibility.}
\label{Shape_eps_g=0.1}
\end{figure}

\begin{enumerate}

\item The stiffness $\tilde \beta$ is taken in the standard form for four-fold anisotropy - such that $\tilde \beta$ is negative for certain step 
orientations $\theta$ when either $\alpha > 1/15$, or $\alpha < -1/15$ (strong anisotropy). In the former (latter) case the stiffness is minimum (maximum) in the direction $\theta=0$
(the $z$-axis). Notice that if the need arises, the general (m-fold) anisotropy can be easily incorporated into the derivation of the evolution equation and its effects are straightforward to analyze along
the lines of Sections III and IV. 
\item The boundary condition (\ref{GT}) for the concentration of adatoms at the step includes the term which provides corner energy regularization \cite{CGP,GDN,S,MS,SGDNV,OL,LLRV}. 
This term is proportional to the positive adjustable parameter $\bar \delta$, which depends on the modulus of the anisotropy strength $\alpha$.
Mathematically, the latter is necessary in order to keep small the cut-off wavenumber $k_c$ for arbitrary anisotropy strength and thus formally remain within the framework of longwave instability theory - note that 
this important feature is absent from the original longwave model of Ref. \cite{GDN}. 
We assume that $\bar \delta$ is a super-linearly increasing function of $|\alpha|$; the form of this function is of no importance for modeling. Thus an increase of 
$|\alpha|$ (which increases $k_c$) is compensated by an increase of $\bar \delta$ (which decreases $k_c$). Physically, the larger is $|\alpha|$ above the critical value 1/15, the more surface orientations are excluded from the equilibrium shape or, in the dynamical situation, the more surface orientations are unstable with respect to short-wavelength perturbations;
the commensurate increase of $\bar \delta$ in response to this allows to keep the short-wavelength instability under control.

\end{enumerate}

\begin{figure}[h]
\vspace{-2.5cm}
\centering
\includegraphics[width=3.0in]{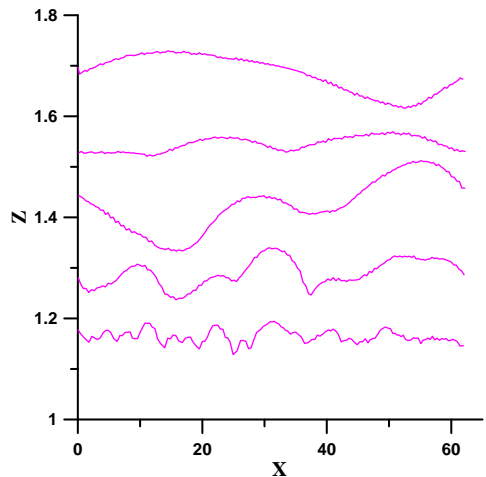}
\vspace{-1.5cm}
\caption{(Color online.) Coarsening of the large wavelength modulation of the mean step position. $\alpha=1.6$, $F=2$. For better view, the profiles corresponding to different times have been bunched together. }
\label{Slow_scale_eps=1.6}
\end{figure}
Due to the four-fold anisotropy of the stiffness,
having negative $\alpha$ values is equivalent to having the corresponding positive values and the phase shift of $\pi/4$ - that is, the stiffness $\tilde \beta =  \beta_0 (1-15\alpha \cos{4\theta})$ with $\alpha <0$
has the same effect on stability and evolution as the stiffness $\tilde \beta =  \beta_0 (1-15|\alpha| \cos{4(\theta-\pi/4)})$. The phase shift is interpreted as rotation of the crystal
that exposes (to instability) a step with another crystallographic orientation. We assume zero phase shift in the expression for $\tilde \beta$, as shown above; non-zero phase shift, if it were factored in, does not have an impact on the main conclusions from the modeling.

Our analysis begins with the formal longwave expansion as in Ref. \cite{GDN}, as follows:
\begin{subequations}
\label{Expansions}
\begin{eqnarray}
x & = & \frac{X}{\epsilon}, t = T_0+\frac{T_2}{\epsilon^2}+\frac{T_3}{\epsilon^3}+\frac{T_4}{\epsilon^4}+...,\\
C & = & C_0(X,z,T_0,T_2,...)+\\
& & \epsilon^2 C_2(X,z,T_0,T_2,...)+...,\nonumber
\end{eqnarray}
\end{subequations} 
where $X$ is the long-scale spatial coordinate, $T_0$ is the fast time, $T_2, T_3, ...$ are the slow time variables and $\epsilon \ll 1$ is the small 
and dimensionless expansion parameter. Physically, $\epsilon$ can be thought of being the ratio of the initial upper terrace width, $H(t=0)$, to the wavelength of the most dangerous (fastest growing) unstable perturbation of the terrace edge (the step). This definition of $\epsilon$ and the scalings (\ref{Expansions}) are different from Refs. \cite{SU} 
and \cite{BMV}, where $\epsilon$ measures the distance from the instability threshold in the weakly nonlinear analysis.
Notice that we do not expand the step position $H$, thus $H(X,T_0,T_2,...)$ is $O(1)$ in $\epsilon$, meaning that the resulting 
longwave evolution equation is strongly nonlinear and thus it is capable of describing deformations of the step of order unity. This is contrasted to the weakly nonlinear equations derived in Refs. \cite{SU,BMV}.

We assume that the density of kinks in the step corner region is high - much higher than a typical density of steps in the surface corner region - and thus the
quantity $\bar \delta = \partial^2E(n,0)/\partial n_x^2$ (which determines the magnitude of the regularization effect) is of the order 
$\epsilon^{-2}$. 
This is responsible for 
a finite interval of unstable wavenumbers already at the order $O(\epsilon^2)$ of the perturbation expansion, as can be seen from Eq. (\ref{EvEq1}) that is derived below.
In the experiments it was determined that the step stiffness is sensitive to ``many-body interactions such as kink-kink interactions and/or effective corner energies" \cite{DIGE}. If by stiffness one understands a full regularized expression, see the comment to Eq. (\ref{GT}), then our physical model qualitatively correlates with these findings.

At the order $O(1)$ we obtain:
\begin{subequations}
\begin{eqnarray}
C_0 & = & \left(C_{eq} -\tau f\right)e^{(H-z)/x_s} + \tau f, \label{C0}\\
H_{T_0} & = & D\Omega \frac{\partial C_0}{\partial z}_{|z=H} = \Omega \sqrt{\tau D}\left(f-f_{eq}\right) \equiv V_0,
\end{eqnarray}
\end{subequations}
where $\tau = x_s^2/D$, and $f_{eq}=C_{eq}/\tau$ is the flux at the equilibrium. Eq. (\ref{C0}) coincides with eq. (8) from Ref. \cite{BMV}. Also, $V_0$ is the constant speed of advance in the positive $z$-direction of a straight (unperturbed) step \cite{SU}, and after transforming to 
the reference frame advancing with this speed, the dependence of $H$ on $T_0$ is eliminated.

At the order $O(\epsilon^2)$ we obtain:
\begin{subequations}
\begin{eqnarray}
C_2 &=& \frac{1}{2}\left(C_{eq} -\tau f\right)e^{(H-z)/x_s}(z-H)\left[\frac{1}{x_s}H_X^2+H_{XX}\right]\pm  \nonumber \\
& & C_{eq}\frac{\Omega \beta_0 }{k_B\bar T}(15\alpha-1)e^{(H-z)/x_s}H_{XX}, \label{CeqOrder2}  \\
H_{T_2} &=& D\Omega \left[\frac{\partial C_2}{\partial z}_{|z=H}-\frac{\partial C_0}{\partial X}_{|z=H}H_X \right] = \nonumber \\
& & x_s^2 \Omega\left[\left\{ \frac{1}{2}\left(f_{eq}-f\right)-f_{eq}\frac{\Omega \beta_0 }{x_s k_B\bar T}(15\alpha-1) \right\}H_{XX} - \right. \nonumber \\
& & \left.f_{eq}\frac{\Omega \delta(|\alpha|) }{x_s k_B\bar T}H_{XXXX} -\frac{f_{eq}-f}{2x_s}H_X^2 \right]. \label{EvEq1}
\end{eqnarray}
\end{subequations}
Notice that at the step, $z=H$, the concentrations are: 
\begin{equation}
C_0 = C_{eq}, \quad C_2=\pm C_{eq}\left(\Omega \beta_0/k_B\bar T\right)(15\alpha-1)H_{XX}. 
\label{C_at_step}
\end{equation}
Also notice that the RHS of Eq. (\ref{GT}) implies $C<C_{eq}$ when $\bar \delta=0$ and $\tilde \beta,\ H_{XX} < 0$ (the step is concave downward). Thus in order to preserve this property - that is, to have $C=C_0+\epsilon^2 C_2=
C_{eq}+\epsilon^2 C_2 < C_{eq}$ - one has to select the positive sign in the expression (\ref{C_at_step}) for $C_2$ at the step (and in Eq. (\ref{CeqOrder2})) when 
$\alpha > 1/15$, and select the negative sign otherwise.


When the anisotropy strength $\alpha \neq 0$, we assumed the corresponding regularization $\delta(|\alpha|)\neq 0$, and thus Eq. (\ref{EvEq1}) has the \emph{form} of the fourth-order Kuramoto-Sivashinsky (KS)
equation (52) from Ref. \cite{BMV}. It incorporates the linear effect of the anisotropy. 
(By \textit{formally} assuming that $\lim_{\alpha \rightarrow 0} \delta(|\alpha|) \neq 0$, the form (52) from Ref. \cite{BMV} is recovered for zero anisotropy.)
Again speaking only about equation structure, Eq. (\ref{EvEq1}) differs from the evolution equation (3.9) in Ref. \cite{SU} in that the anisotropy term in the curly brackets 
features the characteristic combination $15\alpha -1$, while at the same time it is not multiplied neither by the nonlinearity, nor by $\epsilon$ - which makes 
\emph{the linear stability} of the step dependent on anisotropy (similar to the models of surface anisotropy, which Ref. \cite{GDN} pioneered). 
%
\begin{figure}[h]
\centering
\includegraphics[width=2.0in,angle=-90]{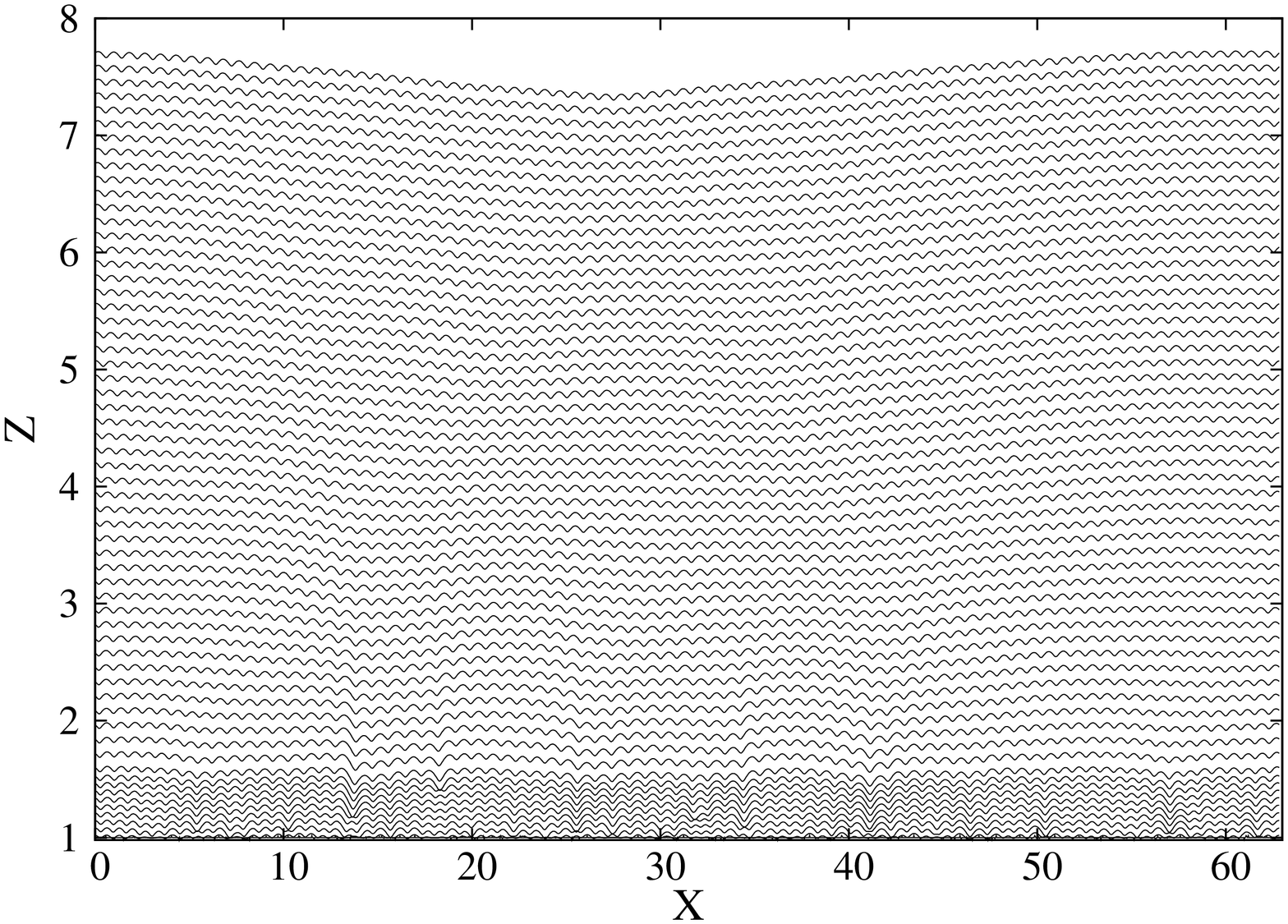}\\ 
\includegraphics[width=2.0in,angle=-90]{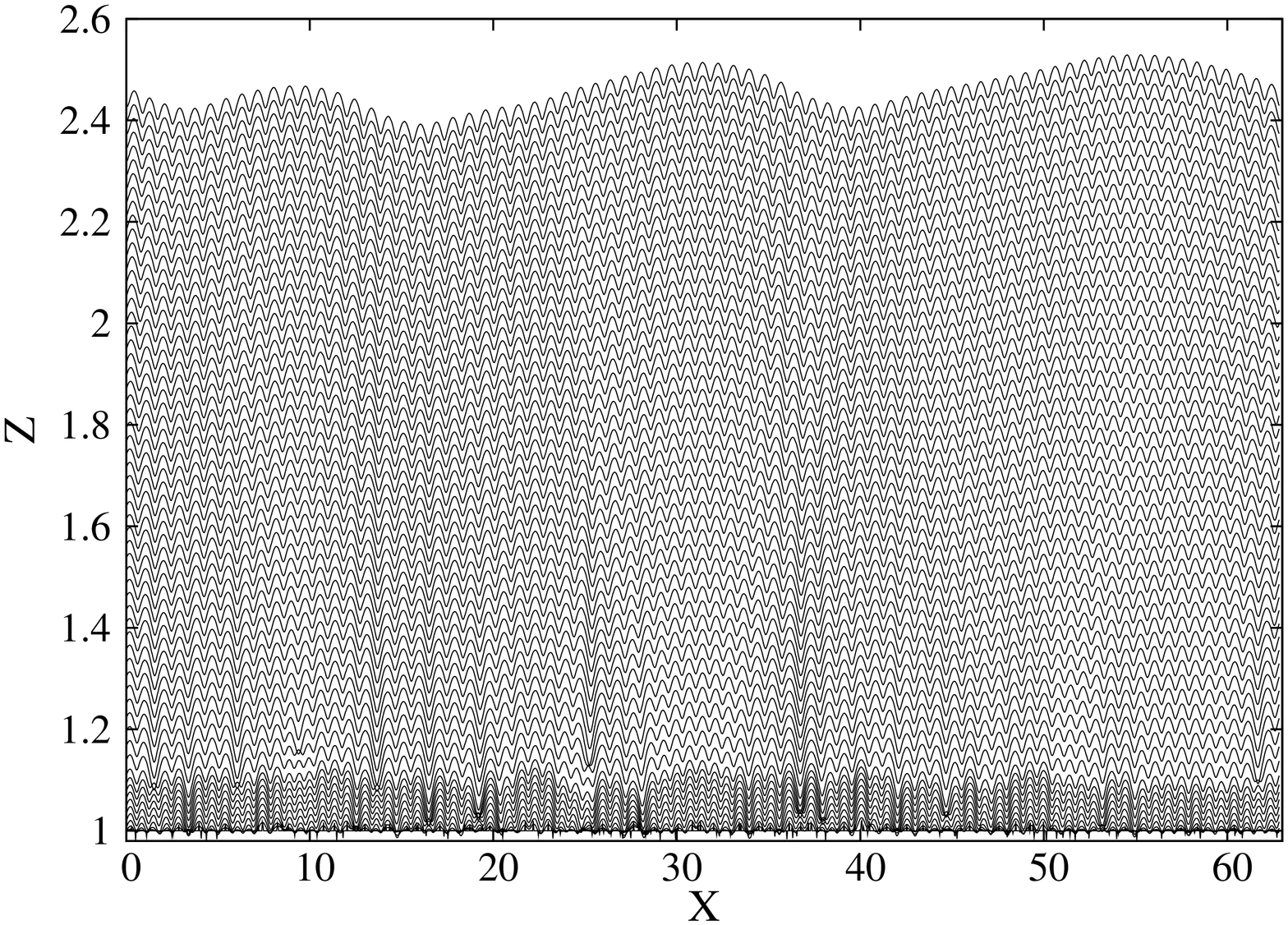}
\caption{Evolution of the morphology from the small random perturbation to the final computed shape. $F=2$. Top panel: $\alpha=0.1$, bottom panel: $\alpha=1.6$. }
\label{AllProfiles_eps_g=0.1_1.6}
\end{figure}
Having obtained the linear effect of the anisotropy, our goal now is to derive the nonlinear contribution.
Thus we proceed to the order $O(\epsilon^3)$ in the perturbaton expansion (where we obtain $H=\mbox{const}(T_3)$) and then to the order $O(\epsilon^4)$, which results in:
\begin{eqnarray}
H_{T_4} &=& D\Omega \left[-\frac{\partial C_2}{\partial X}_{|z=H}H_X -\frac{1}{2}\frac{\partial C_2}{\partial z}_{|z=H}H_X^2+ \right.
\label{H_T4} \\
& & \left. \frac{1}{2}\frac{\partial C_0}{\partial X}_{|z=H}H_X^3\right] = \nonumber \\
& & \frac{x_s^2 \Omega}{4}\left(f_{eq}-f\right)\left[\frac{3}{x_s}H_X^4+H_{XX}H_X^2\right] \mp  \nonumber \\
& & \frac{f_{eq} x_s^2 \Omega^2 \beta_0}{k_B\bar T}(15\alpha-1)\left[\frac{1}{2x_s}H_{XX}H_X^2+H_{XXX}H_X\right].\nonumber
\end{eqnarray}
Notice that all terms at the RHS of Eq. (\ref{H_T4}) are nonlinear and thus they do not affect linear stability. 

Next, combining derivatives like this:
\begin{equation}
H_t = \epsilon^{2}H_{T_2} + \epsilon^{4}H_{T_4},
\label{combination}
\end{equation}
and introducing the original variable $x$ (which cancels the powers of $\epsilon$ in Eq. (\ref{combination})) results in the final, yet dimensional, evolution equation:
\begin{eqnarray}
H_t &=& x_s^2 \Omega\left[\left\{ \frac{1}{2}\left(f_{eq}-f\right)-f_{eq}\frac{\Omega \beta_0 }{x_s k_B\bar T}(15\alpha-1) \right\}H_{xx} - \right.\nonumber\\
& & \left. f_{eq}\frac{\Omega \delta(|\alpha|) }{x_s k_B\bar T}H_{xxxx} -\frac{f_{eq}-f}{2x_s}H_x^2 \right] + \nonumber \\ 
& & \frac{x_s^2 \Omega}{4}\left(f_{eq}-f\right)\left[\frac{3}{x_s}H_x^4+H_{xx}H_x^2\right] \mp \label{EvEq2} \\ 
& & \frac{f_{eq} x_s^2 \Omega^2 \beta_0}{k_B\bar T}(15\alpha-1)\left[\frac{1}{2x_s}H_{xx}H_x^2+H_{xxx}H_x\right].\nonumber
\end{eqnarray}

Lastly, using $x_s$ for the length scale and $\tau$ for the time scale, we obtain the dimensionless evolution equation:
\begin{eqnarray}
H_t & = & \left(m_1-m_2\right)H_{xx} -m_3H_{xxxx} +\frac{m_1 \mp m_2}{2}H_{xx}H_x^2 + \nonumber \\
& & m_1\left(\frac{3}{2}H_x^4-H_x^2\right) \mp m_2H_{xxx}H_x. \label{EvEq5}
\end{eqnarray}
Here: 
\begin{eqnarray}
m_1 & = &\frac{1}{2}\left(f_{eq}-f\right)\Omega \tau,\quad m_2=\frac{f_{eq}\Omega^2 \beta_0 \tau}{k_B \bar T x_s}(15\alpha-1), \nonumber \\
m_3 & = & \frac{f_{eq}\Omega^2 \tau \delta(|\alpha|)}{k_B \bar T x_s^3}.
\label{parameters}
\end{eqnarray}
The parameter $m_1$ measures the deviation of the flux from the equilibrium value, the parameter $m_2$ measures the strength of the anisotropy, and the parameter $m_3$ measures the effect of the regularization (corner rounding). We note the symmetry of Eq. (\ref{EvEq5}) with respect to the transformation $x\rightarrow -x$.
Again comparing to the anisotropy model in Ref. \cite{SU}, Eq. (\ref{EvEq5}) contains the nonlinear terms proportional to $H_{xxx}H_x$ and $H_x^4$ - these terms are not in Eq. (3.9) of Ref.  \cite{SU}.

Since $m_3>0$, clearly, the step is linearly unstable if
\begin{equation}
m_1 - m_2 < 0,\; \mbox{or}\; f > f_c = f_{eq}\left( 1-\frac{2\Omega\beta_0}{x_s k_B\bar T}(15\alpha-1)\right),
\label{flux}
\end{equation}
where $f_c$ is the critical flux of adatoms from the lower terrace to the step.
When $\alpha=0$ (isotropy), the condition (\ref{flux}) coincides with Eq. (2.9) in Ref. \cite{SU}. 

However, we emphasize that Eq. (\ref{EvEq5}) is not applicable in the isotropic case. Indeed, the second and the fifth (the last) 
terms at the RHS of  Eq. (\ref{EvEq5}) are responsible for the onset and development of the step faceting instability, and these terms remain in the equation even 
in the limit $\alpha = 0$ (when there can't be any faceting and/or corner formation). Thus in the isotropic case Eq. (\ref{EvEq1}) where $\alpha$ is set equal to zero and the coefficient of the fourth derivative term is appropriately 
re-defined (as in Ref. \cite{BMV}, for instance) must be used instead. In the weakly anisotropic case ($\alpha\neq 0,\;|\alpha| < 1/15$), because Eq. (\ref{EvEq5}) does not connect
smoothly to the ``isotropic" equation as $\alpha \rightarrow 0$, and because Eq. (\ref{EvEq1}) is deficient (since it does not contain nonlinear terms responsible for faceting and only correctly, as we believe, describes the linear instability), Eq. (3.9) from Ref. \cite{SU} is appropriate for computing the nonlinear dynamics of the step.

\section{Analysis of the evolution equation and its further simplification}
\label{Analysis}

As far as the linear stability of the step governed by Eq. (\ref{EvEq5}) is in question, we observe that the dispersion curve $\omega(k)$ has the typical longwave shape on the interval $0\le k\le k_c$; the growth rate $\omega$ vanishes at the cut-off wave number 
$k_c=\sqrt{(m_2-m_1)/m_3}$. As per our assumptions, $k_c$ remains finite (and small) as $|\alpha|,\; \delta(|\alpha|) \rightarrow \infty$. The most dangerous wavelength
$\lambda_{max} = 2\sqrt{2}\pi/k_c$ and the most dangerous growth rate $\omega_{max} = (m_2-m_1)^2/4m_3$.

Consider the case $\alpha>1/15$ in Eq. (\ref{flux}), i.e. the stiffness is minimum in the direction of growth.
Then, the critical flux is less than the equilibrium one, and at $\alpha = \alpha_c = 1/15+r$, where $r=x_s k_B\bar T/30\Omega\beta_0$ the critical flux vanishes - thus at $\alpha >\alpha_c$ any flux destabilizes the step.  As suggested by the dimensional parameters \cite{Gillet}, $r$  typically is a value between 0.01 and 0.1.
At any flux value such that $f>f_{eq}>f_c$ the growth (in the frame moving with the non-zero velocity $V_0$) and  the instability co-exist \cite{BMV}; at $f_c<f<f_{eq}$, the speed $V_0$ is zero but the step is instable
and growing in the laboratory frame; and at a non-zero $f$, such that $f<f_c<f_{eq}$, the speed $V_0$ is zero and the step is straight, stable and not growing in the laboratory frame. The latter situation is possible only when $\alpha$ is in a narrow interval, $1/15<\alpha<\alpha_c$.


On the other hand, when the stiffness is maximum in the direction of growth ($\alpha<-1/15$), the critical flux needed for instability is larger than $f_{eq}$, signaling that the step is less prone to destabilization. Even when $f< f_{eq}$ and the step is linearly stable,
it is possible to envision the situation when the initial shape of the step is the large-amplitude, smooth (or nearly smooth) curve. For such initial condition 
the nonlinear phase of evolution can be studied; we defer this to future investigations.  
And no matter how $f$, $f_{eq}$ and $m_1$, $m_2$ compare, the nonlinear evolution with $\alpha> 1/15$ 
and with $\alpha<-1/15$ is expected to be different due to the opposite sign of (at least) the last nonlinear term in Eq. (\ref{EvEq5}) (notice how $m_1-m_2$ and $m_1+m_2$ may in principle be of the 
same sign, and then the effective sign of the $H_{xx}H_x^2$ term is the same for $\alpha<-1/15$ and $\alpha>1/15$ cases).

Next, due to the assumption that the regularization parameter $\delta$ is a function of $|\alpha|$, 
the dimensionless parameter $m_3$ can be represented as a function of the dimensionless parameter $m_2$ - thus these parameters are not independent.
Since our goal is to investigate in detail the interplay of the parameters $m_1$ and $m_2$, in the case of linear instability, we eliminate the parameter $m_3$. We do this by fixing the most dangerous wavenumber $k_{max}$ of the linear instability 
and then representing $\delta$ as 
$\delta=(m_2-m_1)k_B\bar T x_s^3/2\Omega^2f_{eq}\tau k_{max}^2$, where $m_2-m_1 >0$. Then the evolution equation takes the form:
\begin{eqnarray}
H_t & = & \left(m_1-m_2\right)\left(H_{xx} + \frac{H_{xxxx}}{2k_{max}^2}\right)+\frac{m_1 \mp m_2}{2}H_{xx}H_x^2 + \nonumber \\
& & m_1\left(\frac{3}{2}H_x^4-H_x^2\right) \mp m_2H_{xxx}H_x, \label{EvEq6}
\end{eqnarray}
where $k_{max}$ is a fixed input parameter.

Before presenting the results of computations, we mention that by introducing the step slope $q \equiv H_x$, Eq. (\ref{EvEq5}) can be put in the following convective (and fully conservative) form \cite{GDN}:
\begin{equation}
q_t + \left(m_1 q^2 - \frac{3}{2}m_1 q^4 - m_2 q_x^2\right)_x + \left(\frac{-\partial G}{\partial q} + m_3 q_{xx}\right)_{xx} = 0,
\label{CH}
\end{equation}
where
\begin{equation}
G = \frac{m_1-m_2}{24}q^4 + \left(\frac{m_1-m_2}{2}-\frac{m_2}{2}q_x\right)q^2.
\label{CH-energy}
\end{equation}
For simplicity of the demonstration, we kept the signs in Eqs. (\ref{CH}), (\ref{CH-energy}) that correspond only to the case $\alpha> 1/15$. The energy $G$ in Eq. (\ref{CH-energy}) is reminiscent of the Cahn-Hilliard energy, although the cubic term is absent and the pre-factor to the quadratic term is the linear function of the step curvature, $q_x$. We found that the computations of Eq. (\ref{CH}) are not as convenient to interpret as those of Eq. (\ref{EvEq6}), thus in the next section we describe
the computations of the latter equation.

\section{Computational results}
\label{Res}

We performed computations of step instability and growth starting from a random, small-amplitude perturbation of the step profile $H(x,0)=1$ on the domain 
$0\le x\le 100\lambda_{max}$ ($\lambda_{max} = 2\pi/k_{max}$), 
with periodic boundary conditions. 
Integration in time of Eq. (\ref{EvEq6}) was performed using the stiff ODE solver DVODE, whereas the discretization in space was carried out using the second order centered finite differencing on a spatially uniform grid.
The number of spatial grid points was at least twenty per wavelength $\lambda_{max}$. Dimensional parameters were chosen as in Ref. \cite{Gillet}. We describe separately the results for the cases $\alpha > 1/15$ and $\alpha < -1/15$.

When the dynamics is governed by a convective Cahn-Hilliard equation, it is known that the pattern length scale coarsens non-uniformly (with different speeds) in time \cite{PZRGN}.
Such is the case, for instance, when the hill-and-valley structure is formed on a crystal surface which evolves by the deposition flux and the surface diffusion with the strongly anisotropic surface energy \cite{SGDNV,OL}.
For the growing step with the strongly anisotropic line energy, absent the line diffusion, we found a similar situation; the computations aim to quantify the growth speed and the laws of the morphology coarsening.
\begin{figure}[h]
\vspace{-2.0cm}
\centering
\includegraphics[width=3.0in]{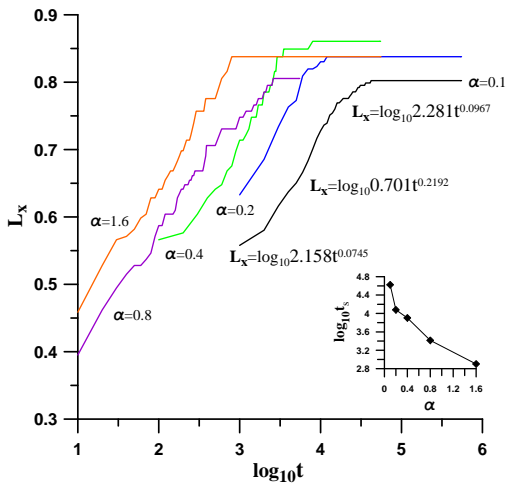}
\vspace{-1.5cm}
\caption{(Color online.) Coarsening of the hill-and-valley structure for different anisotropy strengths. Coarsening laws are shown for the case $\alpha=0.1$. Inset shows the time to reach the steady-state 
(the horizontal part of the curve in the main figure) vs. $\alpha$. (The curve in the inset is only the guide for the eye.) $F=2$.}
\label{Coarsening_F=2_alpha_varies}
\end{figure}

\subsection{Stiffness is minimum in the growth direction ($\alpha > 1/15$)}
\label{StiffMin}

In this section we describe the results of the two sets of computations: the first, with the fixed flux $f=2f_{eq}$ and $\alpha = 0.1,0.2,0.4,0.8,1.6$; and the second, with fixed $\alpha=0.4$ and 
$f/f_{eq}\equiv F =0.1,1,10,50,150$. For these values of $\alpha$ the critical flux is zero (see the above analysis) thus for all chosen $f$ values the step is linearly unstable.

\subsubsection{Fixed flux}
\label{FF}

Fig. \ref{Shape_eps_g=0.1} shows the step morphology after the final length scale of the hill-and-valley structure has been established - that is, hills and valleys do not coarsen any more; we call this a steady-state. Such late-time morphology is typical to all anisotropy strengths $|\alpha|>1/15$.  The length scale of the hill-and-valley structure is defined as the ratio of the length of the computational domain to the number of kinks (valleys).
Besides this length scale one can clearly distinguish a large-amplitude, large wavelength modulation of the mean step position (the dashed line). The latter length scale continues to coarsen even after the steady-state
has been reached. For example, we show the coarsening of the large wavelength modulation in Fig. \ref{Slow_scale_eps=1.6}, but at this point we are unable to quantify it due to difficulties of the numerical implementation. 
Also note, from the top inset of Fig. \ref{Shape_eps_g=0.1}, that the slope of the hills is fairly gentle - around $12^\circ$. 

Fig. \ref{AllProfiles_eps_g=0.1_1.6} shows the entire time evolution of the morphology for $\alpha=0.1$ and $\alpha=1.6$. Especially in the right panel one can notice that 
the long-wavelength modulation continues to coarsen even after the steady-state has been reached roughly by the time it takes 
the step to grow to the level $z=1.6$. Prior to that, the significant changes in the length scale of the hill-and-valley structure (the kink-antikink collisions) are signaled
by the regions of high black contrast.

Fig. \ref{Coarsening_F=2_alpha_varies} shows the plots of the length scale of the hill-and-valley structure, $L_x$, vs. the time for varying anisotropy strength $\alpha$. ($L_x$ was averaged over ten computations with the different initial random perturbation of the straight step.) 
One can distinguish three coarsening regimes for $\alpha = 0.1, 0.2$ and $0.4$, and two regimes for $\alpha = 0.8$ and $1.6$.
Simple averaging gives $\langle L_x\rangle = \log_{10}{2.65t^{0.054}},\; \langle L_x\rangle = \log_{10}{1.27t^{0.203}},\; \langle L_x\rangle = \log_{10}{1.88t^{0.154}}$ for the first, second, and third regime, respectively.  Clearly, coarsening is the fastest in the second regime.
As has been pointed above, the existence of different coarsening regimes for given $\alpha$ can be 
attributed to the presence of convective terms in the Cahn-Hilliard-like evolution equation (\ref{CH}) \cite{PZRGN}.
The final, steady-state value of $L_x$ is not very sensitive to $\alpha$; this value (in the 0.8-0.9 range) seems to depend on $\alpha$ 
non-monotonically. The value signals that the final number of kinks (or anti-kinks) is significantly less (by 20-25\%) than one hundred - the latter number is the number of kinks that would fit 
into the computational domain according to the linear stability analysis.
Also notice that as the anisotropy strength increases, the steady-state occurs faster (inset). 
\begin{figure}[h]
\vspace{-2.0cm}
\centering
\includegraphics[width=3.0in]{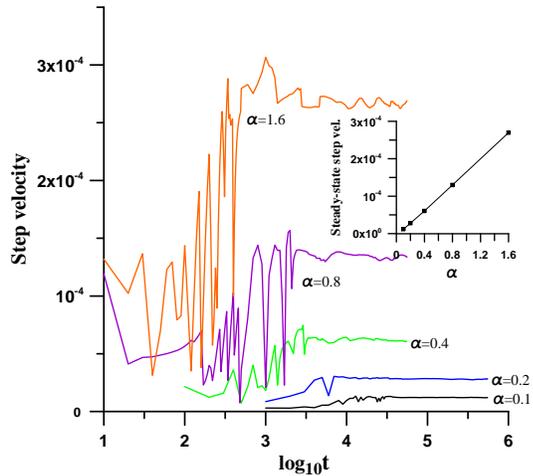}
\vspace{-1.5cm}
\caption{(Color online.) Step velocity vs. the time for different anisotropy strengths. Inset shows the velocity in the steady-state 
(the horizontal part of the curve in the main figure) vs. $\alpha$. The slope of this line is $1.7\times 10^{-4}$. (The curve in the inset is only the guide for the eye.) $F=2$.}
\label{Velocity_F=2_alpha_varies}
\end{figure}

Finally, Fig. \ref{Velocity_F=2_alpha_varies} shows the graphs of the step velocity vs. the time for different anisotropy strengths. Irregularity (oscillation) of the velocity increases with $\alpha$, but ultimately a steady-state velocity is reached.  Comparison of Fig. \ref{Coarsening_F=2_alpha_varies}
with Fig. \ref{Velocity_F=2_alpha_varies} shows that the emergence of the steady-state velocity coincides with the establishment of the final length scale of the hill-and-valley structure. The inset to this Figure demonstrates that the steady-state velocity increases linearly with the anisotropy strength.

\subsubsection{Fixed anisotropy strength}
\label{FAS}

\begin{figure}[h]
\vspace{-2.0cm}
\centering
\includegraphics[width=3.0in]{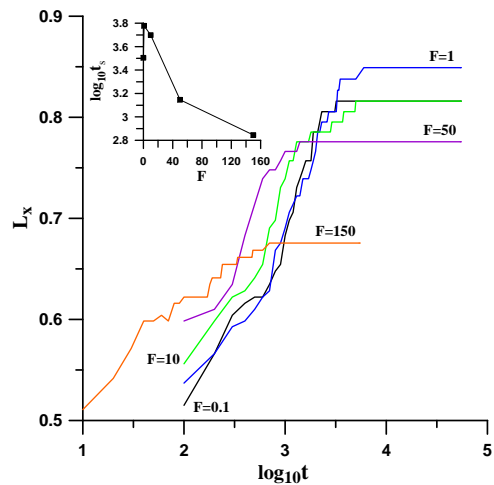}
\vspace{-1.5cm}
\caption{(Color online.) Coarsening of the hill-and-valley structure for different fluxes $F$.  Inset shows the time to reach the steady-state 
vs. $F$. (The curve in the inset is only the guide for the eye.) $\alpha=0.4$.}
\label{Coarsening_alpha=0.4_F_varies}
\end{figure}

Fig. \ref{Coarsening_alpha=0.4_F_varies} shows the plots of the length scale, $L_x$, vs. the time for the varying flux $F$.   
In contrast to Fig. \ref{Coarsening_F=2_alpha_varies} we notice that both the time to the steady-state (inset) and the steady-state length scale are the non-monotonic functions of the flux.
Both increase when $F\sim 1$, then decrease. When $F$ is increased from 0.1 to 150, the length scale decreases by 20\%.  For flux values larger than 200 the step evolution is chaotic.

\begin{figure}[h]
\vspace{-2.0cm}
\centering
\includegraphics[width=3.0in]{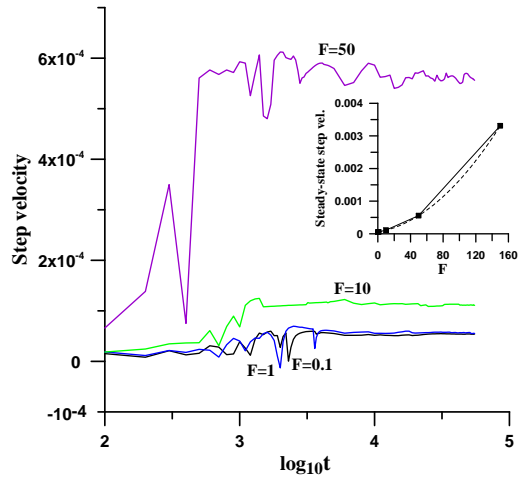}
\vspace{-1.5cm}
\caption{(Color online.) Step velocity vs. the time for different fluxes $F$. The curve that corresponds to $F=150$ is not shown in order to not scale down the view of the other four curves. 
Inset shows the velocity in the steady-state. 
(The solid curve in the inset is only the guide for the eye.) The dashed curve is the five-points fit $1.16\times 10^{-7} F^2 + 4.26\times 10^{-6} F + 5.35\times 10^{-5}$. $\alpha=0.4$.}
\label{Velocity_alpha=0.4_F_varies}
\end{figure}

Fig. \ref{Velocity_alpha=0.4_F_varies} shows the plots of the step velocity vs. the time for the varying flux $F$. Predictably, velocity increases when the flux increases.
As is shown in the inset, this dependence is nonlinear; it is fitted quite well by the quadratic polynomial.


\subsection{Stiffness is maximum in the growth direction ($\alpha < -1/15$)}
\label{StiffMax}

Here, because the problem is very numerically stiff even with regularization in effect, we succeeded to compute the step evolution only for modest values of the anisotropy strength $|\alpha|$ above the 
critical value $1/15$: $1/15 < |\alpha| \le 0.2$. Since when $\alpha < -1/15$ the critical flux is 
always non-zero (see Eq. (\ref{flux}) and its discussion), then every time a new $\alpha$ is chosen, the input flux must also be changed to keep it above $f_c$. The values we chose are: 
for $\alpha=-0.1$, $F=15.5$;  for $\alpha=-0.15$, $F=18.8$; $\alpha=-0.2$, $F=22.2$. These values are at the same distance ($=3.3$, arbitrarily chosen) from the corresponding $F_c$ value in all three cases.

\begin{figure}[h]
\vspace{-2.0cm}
\centering
\includegraphics[width=3.0in]{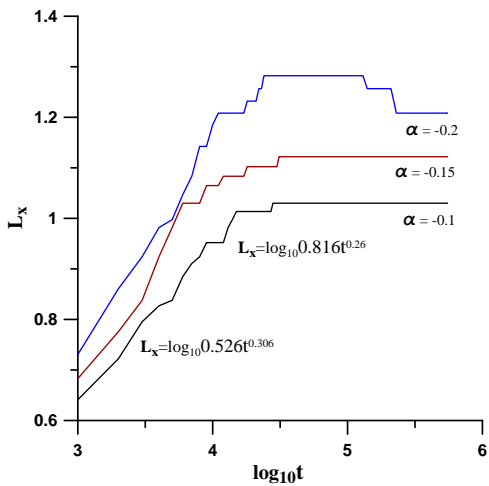}
\vspace{-1.5cm}
\caption{(Color online.) Coarsening of the hill-and-valley structure for different anisotropy strengths. Coarsening laws are shown for the case $\alpha=-0.1$.}
\label{Coarsening_F_varies_alpha_varies}
\end{figure}

Fig. \ref{Coarsening_F_varies_alpha_varies} shows the plots of the length scale, $L_x$, vs. the time for the varying $\alpha$. Notice, by comparison with Fig. \ref{Coarsening_F=2_alpha_varies},
that the coarsening exponents are much larger in the present case, and the steady-state length scales are also much larger, signaling the reduction in the number of kinks compared to the number expected from $\lambda_{max}$ by as much as 50\% even for these relatively small values of $|\alpha|$.
Again averaging the data from the three curves gives $\langle L_x\rangle = \log_{10}{0.47t^{0.342}},\; \langle L_x\rangle = \log_{10}{0.79t^{0.288}}$ for the first and the second regime, respectively.  Coarsening is the fastest in the first regime.

The step velocity exhibits the behavior very similar to the one in Fig. \ref{Velocity_F=2_alpha_varies}, including the linear dependence of the steady-state velocity on the anisotropy strength.

\section{Summary}
\label{Conclude}

In this paper, a longwave PDE is formulated for the description of the strongly anisotropic step dynamics within the framework of a one-sided model.
Written in terms of the step slope, the PDE can be represented in the form similar to a convective Cahn-Hilliard equation. 

Analysis of the model shows, most importantly, that the linear stability of a step depends not only on the 
strength of the adatoms flux from the terrace to the step, but also on the sign and the strength of the line energy anisotropy parameter $\alpha$. 
The latter observation is entirely new. However, it should not come as a complete surprise, since the mathematical structure of the 
one-sided step evolution model is similar to the model of solidification into a hypercooled melt \cite{GDN}. 
That model involves the concentration field dynamics on one side of the 
interface only (in the liquid phase), and it is well-known and accepted that the linear stability 
depends on the anisotropy strength. Consequently, the results of our modeling remind those obtained in Ref. \cite{GDN}.

Specifically, we found that it matters whether the step stiffness is a minimum  or a maximum in the direction of the step growth. In the former case, when $\alpha$ is larger than 
the threshold value, the critical flux that destabilizes the step is \emph{less} than the equilibrium value, and it is even possible to destabilize the step by anisotropy alone by taking 
$\alpha$ large enough. That is, the flux and the anisotropy complement each other in destabilizing the step. 
Though in the latter case, the critical flux is larger than the equilibrium value for any $\alpha$. 

In the computations of the nonlinear dynamics with the strong anisotropy, most interestingly, we found the emergence and coarsening of the long-wavelength deformation of the step,
which goes on simultaneously with the (non-uniform in time) coarsening of the hill-and-valley structure. We characterized the coarsening of the latter structure as a function of the anisotropy and 
the flux, and our future research will focus on the analysis of the former process.

\bigskip
\noindent
{\bf Acknowledgements}

 I am very grateful to Olivier Pierre-Louis for reading the first draft of the paper and making many valuable comments and suggestions. 
Michel Jabbour is acknowledged for bringing the anisotropic step dynamics to my attention.


\begin{thebibliography}{200}

\bibitem{OSYP}   M. Ondrejcek, W. Swiech, G. Yang, and C.P. Flynn, 
\textit{J. Vac. Sci. Technol. B}$\;$ {\bf 20}, 2473 (2002).

\bibitem{OO} H. Omi and T. Ogino, 
\textit{Thin Solid Films}$\;$ {\bf 380}, 15 (2000).

\bibitem{XSZZWT} M. H. Xie, S. M. Seutter, W. K. Zhu, L. X. Zheng, Huasheng Wu, and S. Y. Tong, 
\textit{Phys. Rev. Lett.}$\;$ {\bf 82}, 2749 (1999).

\bibitem{DPKM} G. Danker, O. Pierre-Louis, K. Kassner, and C. Misbah, 
\textit{Phys. Rev. E}$\;$ {\bf 68}, 020601 (2003).

\bibitem{SU} Y. Saito and M. Uwaha, 
\textit{J. Phys. Soc. Jpn.}$\;$ {\bf 65}, 3576 (1996).

\bibitem{MSOW} H. Mori, T. Soma, K. Okuda, and K. Wada, 
\textit{J. Phys. Soc. Jpn.}$\;$ {\bf 73}, 1362 (2004).

\bibitem{S} B.J. Spencer, 
\textit{Phys. Rev. E}$\;$ {\bf 69}, 011603 (2004).

\bibitem{Regular} Expansion of $E\left(n,n_x\right)$ in powers of $n_x$ results in the additive contribution to the surface energy. This additive contribution is proportional to the square of the surface curvature; the coefficient, usually denoted by $\delta$, is called the regularization parameter (\cite{GDN}, p. 208). Accordingly, a corner on a surface (large curvature) posesses an additional energy and thus it is 
penalized - since the system seeks the state of a minimum energy. As pointed out in Refs. \cite{S,LLRV}, the idea of regularization by curvature goes back to the 
classical work of Conyers Herring \cite{Herring}.

\bibitem{Herring} C. Herring, \textit{Phys. Rev.}$\;$ {\bf 82}, 87 (1951).

\bibitem{GDN} A.A. Golovin, S.H. Davis, and A.A. Nepomnyashchy, 
\textit{Physica D}$\;$ {\bf 122}, 202 (1998).

\bibitem{CGP} A.\ Di Carlo, M.E.\ Gurtin, and P.\ Podio-Guidugli, 
\textit{SIAM J. Appl. Math.}$\;$ {\bf 52}, 1111 (1992).

\bibitem{MS} A. Mastroberardino and B.J. Spencer, 
\textit{IMA J. Appl. Math.}$\;$ {\bf 75}, 190 (2010).

\bibitem{SGDNV} T.V.\ Savina, A.A.\ Golovin, S.H.\ Davis, A.A.\ Nepomnyashchy, and P.W.\ Voorhees, 
\textit{Phys. Rev. E}$\;${\bf 67},  021606 (2003).

\bibitem{OL} C. Ograin and J. Lowengrub, 
\textit{Phys. Rev. E}$\;$ {\bf 84},  061606 (2011).

\bibitem{LLRV} B. Li, J. Lowengrub, A. Ratz, and A. Voigt, 
\textit{Comm. Comp. Phys.}$\;$ {\bf 6},  433 (2009).

\bibitem{BMV} I. Bena, C. Misbah, and A. Valance, 
\textit{Phys. Rev. B}$\;$ {\bf 47}, 7408 (1993).

\bibitem{DIGE} S. Dieluweit, H. Ibach, M. Giesen, and T.L. Einstein, 
\textit{Phys. Rev. B}$\;$ {\bf 67}, 121410(R) (2003).

\bibitem{Gillet} F. Gillet, O. Pierre-Louis, and C. Misbah, 
\textit{Eur. Phys. J. B}$\;$ {\bf 18}, 519 (2000).

\bibitem{PZRGN} A. Podolny, M.A. Zaks,  B.Y. Rubinstein, A.A. Golovin, and A.A. Nepomnyashchy,  
\textit{Physica D}$\;$ {\bf 201}, 291 (2005).

\end{thebibliography}
\end{document}